\documentclass[final,10pt,twocolumn]{elsarticle}
\usepackage[utf8]{inputenc}
\usepackage{amssymb}
\usepackage{mathtools}
\usepackage{graphicx} 
\usepackage{lineno}
\usepackage[hidelinks]{hyperref}
\usepackage{natbib}

\journal{Journal of non-crystalline solids}

\begin{document}
\begin{frontmatter}


\title{The role of Ce$^{3+}$/Ce$^{4+}$ in the spectroscopic properties of cerium oxide doped zinc-tellurite glasses prepared under air}


\author{Maiara Mitiko Taniguchi$^a$}
\author{Edenilson da Silva$^a$}
\author{Marco Aurélio Toledo da Silva$^b$}
\author{Leandro Herculano$^c$}
\author{Robson Ferrari Muniz$^d$}
\author{Marcelo Sandrini$^a$}
\author{Marcos Paulo Belan\c{c}on$^a$} 
\ead{marcosbelancon@utfpr.edu.br}

 \address[$^a$]{Universidade Tecnol\'{o}gica Federal do Paran\'{a} (UTFPR), Câmpus Pato Branco}
\address[$^b$]{Universidade Tecnol\'{o}gica Federal do Paran\'{a} (UTFPR), Câmpus Londrina, Postgraduate course in Materials Science and Engineering, Laboratory of Photonics and Nanostructured Materials (DFMNano)}
\address[$^c$]{Universidade Tecnol\'{o}gica Federal do Paran\'{a} (UTFPR), Câmpus Medianeira, Departamento de F\'isica}
\address[$^d$]{Universidade Estadual de Maringá (UEM), Campus Goioerê, Departamento de Ciências}

\begin{abstract}
Emerging technologies are demanding innovative properties of glasses. In this work Cerium Oxide is used as dopant in Zinc-Tellurite samples and its effects on the properties of the glass are investigated. Thermal analysis and x-ray diffraction confirmed the amorphous nature of all samples. The bivalent nature of Cerium is investigated spectroscopically and an strong redshift induced by the dopant is attributed to charge transfers O$^{2-}{\rightarrow}$Ce$^{4+}$, while the 4f-5d transition of Ce$^{3+}$ could not be identified in absorption or luminescence measurements. Yellow/red (570/650 nm) emission under 405/450 nm pumping were observed and are originated from Te$^{4+}$ ions, which absorbs light in the UV/blue region of the spectrum. The incorporation of some Cerium Oxide enhanced the visible luminescence, though we found no evidence that Cerium ions play some role in the radiative process. The luminescence is enhanced though due changes in the glass network induced by the dopant.
\end{abstract}

\begin{keyword}
tellurite glass\sep cerium doping\sep active tellurium \sep zinc-tellurite\sep 


\end{keyword}

\end{frontmatter}


\section{Introduction}
\label{sec:intro}

The need to upscale the electricity production from solar and other emerging technologies are pushing the glass science to innovate further \cite{Mauro2014}. Taking just the example of the photovoltaic (PV) industry, about 150 GWp \cite{ITRPV2018} of capacity was installed last year, corresponding to about 500 million panels; virtually all of them are covered by float glass. PV's are already pushing the glass industry \cite{Burrows2015} to increase production, and emerging technologies \cite{Nayak2019} in this field and others are demanding enhanced glass materials.

Rare-earth's (RE) have a straight relation with glasses at the industry level. By mass, cerium oxide is the most consumed RE and it is widely used as UV blocking agent in PV's cover \cite{Oliveira2018}. Considering an average thickness of 3 mm for the cover glass, one can estimate that PV's industry alone consumed more than 5 million tons of float glass last year. Such enhanced glass materials we are pursuing worldwide could reduce such massive raw materials consumption while bringing new functions, such as spectral conversion to these components.

On the other hand, tellurite glasses are interesting hosts due to their unique structure, high refractive index and low melting temperature \cite{Jha2012a}. Tellurites can be fiberized \cite{Belancon2013b}, sputtered \cite{Ogbuu2015} and other techniques as laser ablation of the glass have been developed \cite{Mann2018}. The incorporation of Cerium in this family of glasses has often the purpose to introduce Ce$^{3+}$ ions to quench the $^4$I$_{11/2}$ level of Er$^{3+}$ \cite{Yang2004a,Sasikala2012,Zheng2013} which enhances the gain in optical amplifiers operating at 1.53$\mu m$. However, Ce$^{3+}$ also exhibits a 4f-5d transition, which is very sensitive to the crystal field of the host. This transition introduces intense optical absorption, which can be found between the UV \cite{Stroud1961a} and the orange \cite{Zhang2015b} part of the visible spectrum. These properties open a wide field of applications for Ce$^{3+}$, such as a sensitizer for downconversion of UV/Blue in visible/near-infrared light \cite{Zhou2017}, in order to enhance the efficiency of solar cells. 

Most of the literature on the doping of tellurite glasses with cerium oxide are focused in NIR properties of Er$^{3+}$, as we mentioned, though a clear red-shift of the absorption edge in these glasses is often observed and attributed to the 4f-5d transition of Ce$^{3+}$ \cite{Sasikala2012, Zheng2013, Tao2012, Huang2016a, Su2018}. The radiative emission from decays of the 5d manifold in this ion is widely explored since the 1960's \cite{Knoll1967}, though to the best of our knowledge it has not been reported in tellurite glasses yet. However, Ce$^{4+}$ may also introduce intense UV/VIS absorption bands due O$^{2-}{\rightarrow}$Ce$^{4+}$ charge transfer (CT). In this way, we found it important to investigate the presence of these different valence states of cerium in tellurite glasses, which could potentially be explored to produce spectrally selective windows and coatings to enhance solar cells and other optical devices.

The reduction/oxidation balance of Cerium valence states have been studied for several decades in glasses, and it is known that the ratio between Ce$^{3+}$/$Ce^{4+}$ is sensitive to host composition, melting temperature, melting atmosphere and may even change due to the incidence of light \cite{Stroud1961a, Lin1995, Ebendorff-Heidepriem2000a, Pinet2006, Zhao2018}. Such bivalent presence of Cerium in tellurite glasses still needs a better understanding and in this work, we report some spectroscopic investigation we have performed in a Zinc-Tellurite glass system. This family of glasses has been extensively investigated since the 1980's \cite{Mann2018, Technology1986, Amjad2015, Huang2016, Suresh2019}, and adjusting the concentration of network modifiers, glasses with good thermal stability, transparency and high RE solubility can be obtained. These characteristics highlight the potential of the material and the need to evaluate its behavior when Cerium oxide is incorporated, a combination that could be useful to the development of many optical devices.

\section{Materials and methods}
Samples containing 73.3TeO$_2$, 19.6ZnO, 4.9Na$_2$CO$_3$ and 2.2La$_2$O$_3$ (TZNL) in \%mol were prepared by weighting ($\pm0.1$ mg) the high purity Tellurium Dioxide (99.995\%), Zinc Oxide ($\geqslant$ 99.99\%), Sodium Carbonate ($\geqslant$ 99.5\%) and Lanthanum Oxide (99.999\%) to achieve 10 g glass samples, which were named as xCe, where x is the molar concentration of Cerium IV oxide (99.9\%) added to the TZNL sample. All these raw materials were used as purchased from Sigma Aldrich and the dopant was added to the composition without changing the stoichiometric relation between the components of the TZNL matrix. Next, the raw materials were melted at 800$^o$C in the air atmosphere and poured into stainless steel mold pre-heated to 300$^{o}$C, just below the glass transition temperature. 

The differential scanning calorimetry (DSC) data were acquired with a TA Instruments DSC Q20 using Air atmosphere, with a flow rate of 50 mL/min and a heating rate of 10$^o$C/min. Samples were milled in order to have the same conditions (contact area) for all samples. A platinum crucible was used and the samples were heated until 520$^{o}$C. X-ray diffraction (XRD) were obtained from the glass powder
using a Rigaku Miniflex 600 diffractometer with CuK$\alpha$ radiation of wavelength 1.5418 Å with 30 kV and 15mA current. The scan range was set to 2$\theta$ from 5$^o$ to 80$^o$ with a step size of 0.02$^o$.  UV-VIS absorbance was measured in a PerkinElmer Lambda 45 spectrometer. Visible and near-infrared photoluminescence (PL) spectra were obtained with a LaserLine SP-2A portable spectrometer. A laser and a light emission diode (LED) coupled with a monochromator were used as the light source. Excitation spectra were obtained monitoring the emission at 565 nm, using a Xe lamp and a Newport monochromator model 77780 as an excitation source. The emission intensity was collected by a photomultiplier tube, and a lock-in amplifier (Stanford Research System, model SR830) was used to reduce the noise.

\section{Results}

\subsection{DSC}
\begin{figure}[h]
\centering
\includegraphics[trim=2cm 1cm 2cm 1cm, clip,scale=0.32]{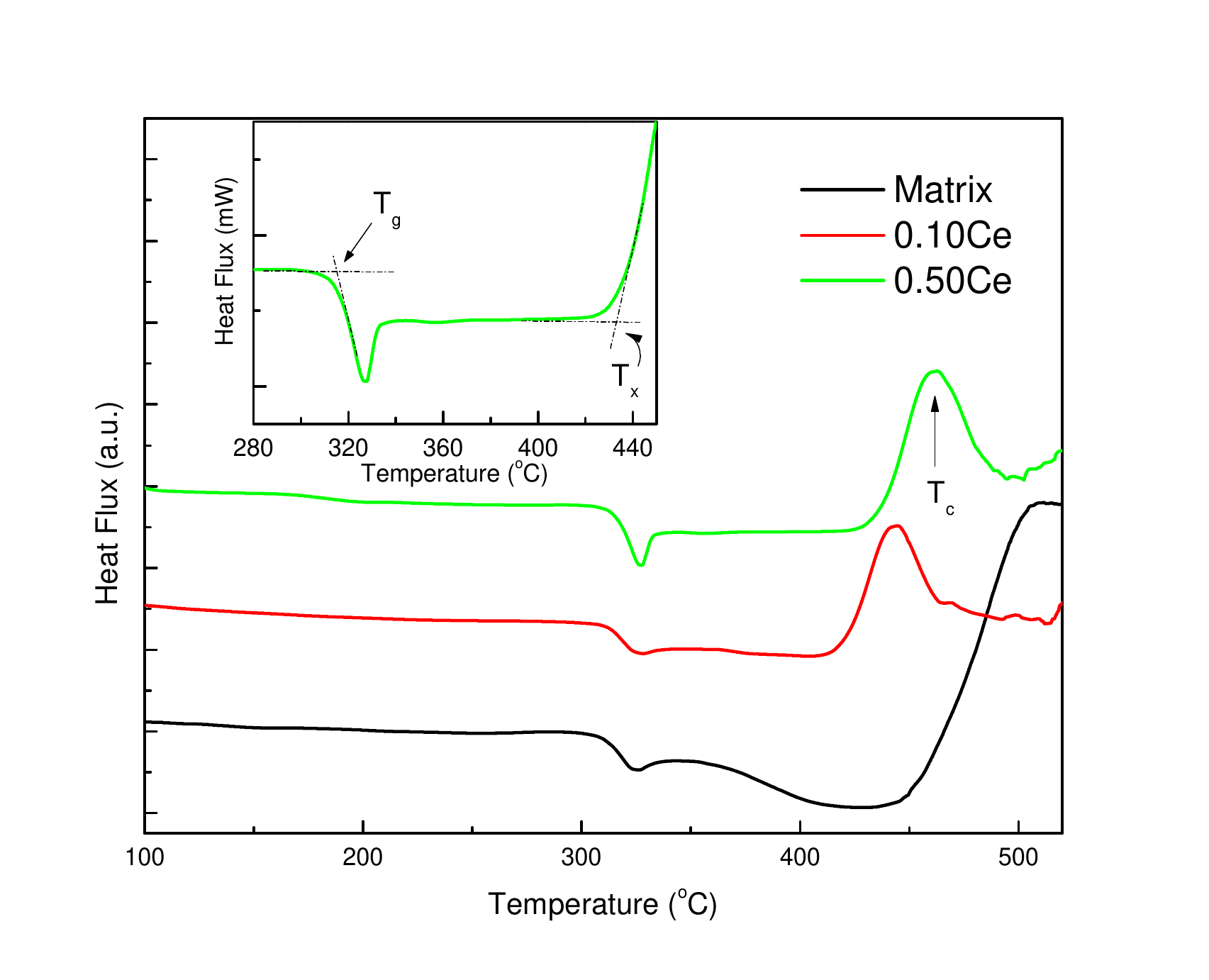}%
\caption{DSC curves evidencing the crystallization peak around 450$^oC$ in doped samples.}
\label{dsc}
\end{figure}

In figure \ref{dsc} we show the DSC curves in the range 100-520$^o$C and the temperatures of glass transition (T$_g$), onset crystallization (T$_x$) and crystallization peak (T$_c$) obtained are summarized in table \ref{dsctable}. Based on the precision of the equipment and in how the samples were prepared, the error estimates for these temperatures are $\pm0.1^o$C. 

\begin{table}[h]
\centering
\begin{tabular}{|c|c|c|c|c|}
\hline
\textbf{Sample}&\textbf{T$_g$}&\textbf{T$_x$}&\textbf{T$_c$}&\textbf{T$_x$-T$_g$}\\
\hline
Matrix&309.1&451.9&506.2&142.8\\
\hline
0.10Ce&313.8&417.0&444.2&103.2\\
\hline
0.50Ce&315.8&433.8&461.7&118.0\\
\hline
\end{tabular}
\caption{T$_g$,T$_x$ and T$_c$ obtained from the curves shown in figure \ref{dsc}.}
\label{dsctable}
\end{table}
As one can see, the T$_g$ has increased with increasing Cerium concentration. This increase is related to changes in the network connectivity caused by Cerium Oxide insertion, i. e. possible modifications on tellurium activated units, from Te$O_3$ and Te$O_{3+1}$ to Te$O_4$. Crystallization also takes place at different temperatures. The shift to lower temperature suggests a premature phase formation, indicating that Ce ions act as specks to stimulate crystallization. The ability of a glass to avoid crystallization upon cooling of supercooled liquids can be analyzed by the difference between T$_x$ and T$_g$. Doped samples exhibited lower T$_x$-T$_g$, though in all cases such values remained above 100$^o$C, which indicates quite good stability.

\subsection{XRD}
\begin{figure}[h]
\centering
\includegraphics[trim=3cm 1cm 2cm 1cm, clip,scale=0.4]{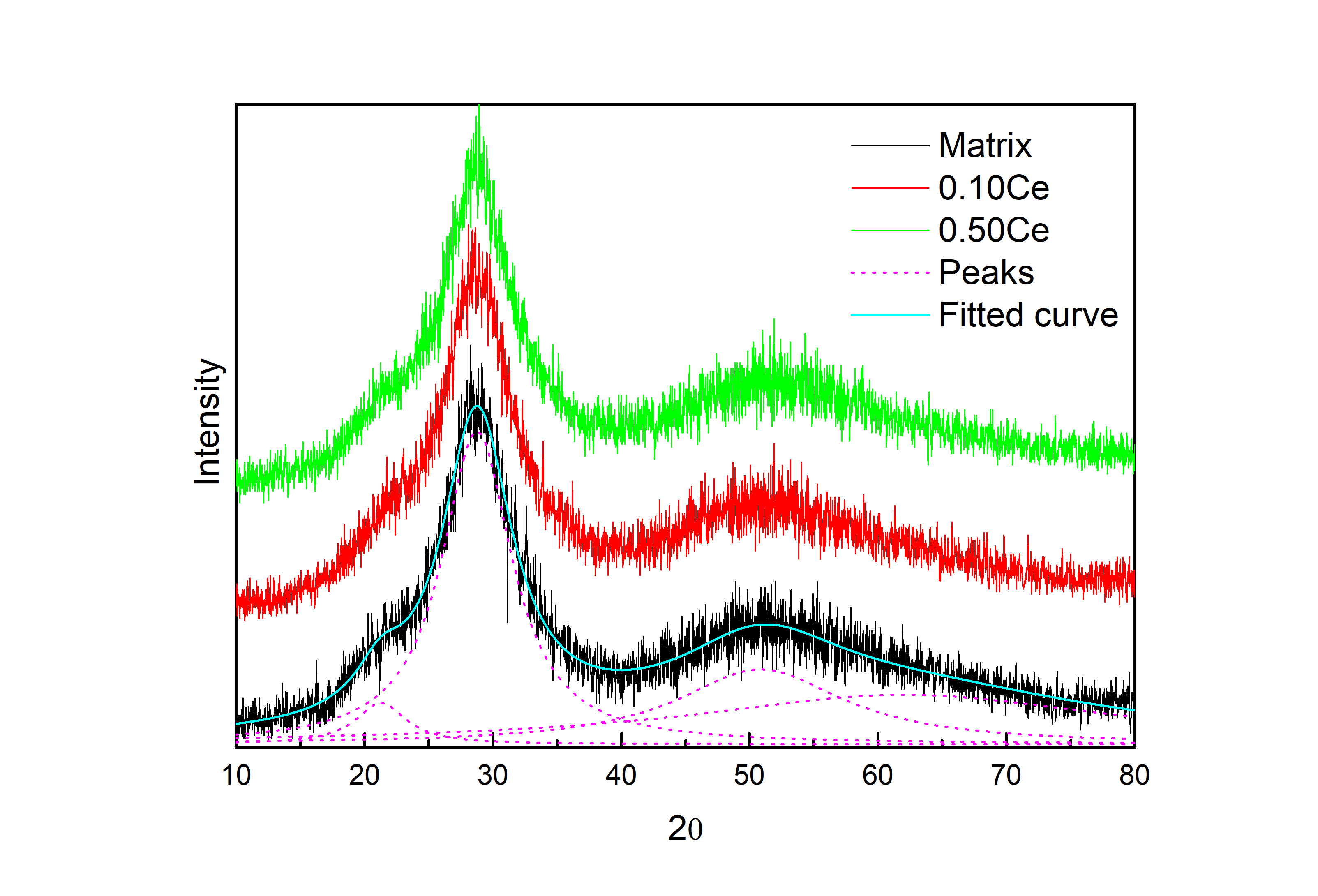}%
\caption{X-ray diffraction profiles and one example of the fit obtained with three lorentzians.}
\label{drx}
\end{figure}

In figure \ref{drx} we can see the X-ray diffractograms for studied samples, which are similar to those obtained by Sobczyk et al \cite{Sobczyk2018} in a similar glass, including for a crystallized 0.10Ce sample (diffractogram not shown here) that was annealed for 11 hours at 470$^oC$. As they have explained, ZnTeO$_3$ and La$_2$Te$_4$O$_{11}$ crystals are formed above 450$^o$C, and the concentration ratio between them is sensitive to the temperature. This could explain why T$_x$ is lower in our doped samples, once Cerium may be changing this balance or introducing a new crystal phase. 
To better identify the peak position observed, we have fitted the diffractograms by four Lorentzians obtaining a chi$^2$ of about 0.0012 in all cases, which resulted in fitted curves as shown in cian in the figure \ref{drx}. The most intense band is centered at 28.7$^o$ for all samples. This peak position is sensitive to the concentration ratio of Zinc/Tellurium in the glass \cite{Tagiara2017}, and as we kept the stoichiometric relation of the matrix in all samples (the dopant was added) such result was expected and indicates that these proportions are maintained in the final samples. The XRD results also confirm that cerium oxide is not affecting the amorphous nature of the samples.

\subsection{UV-VIS absorbance}
\begin{figure}[h]
\centering
\includegraphics[trim=3cm 1cm 2cm 2cm, clip,scale=0.38]{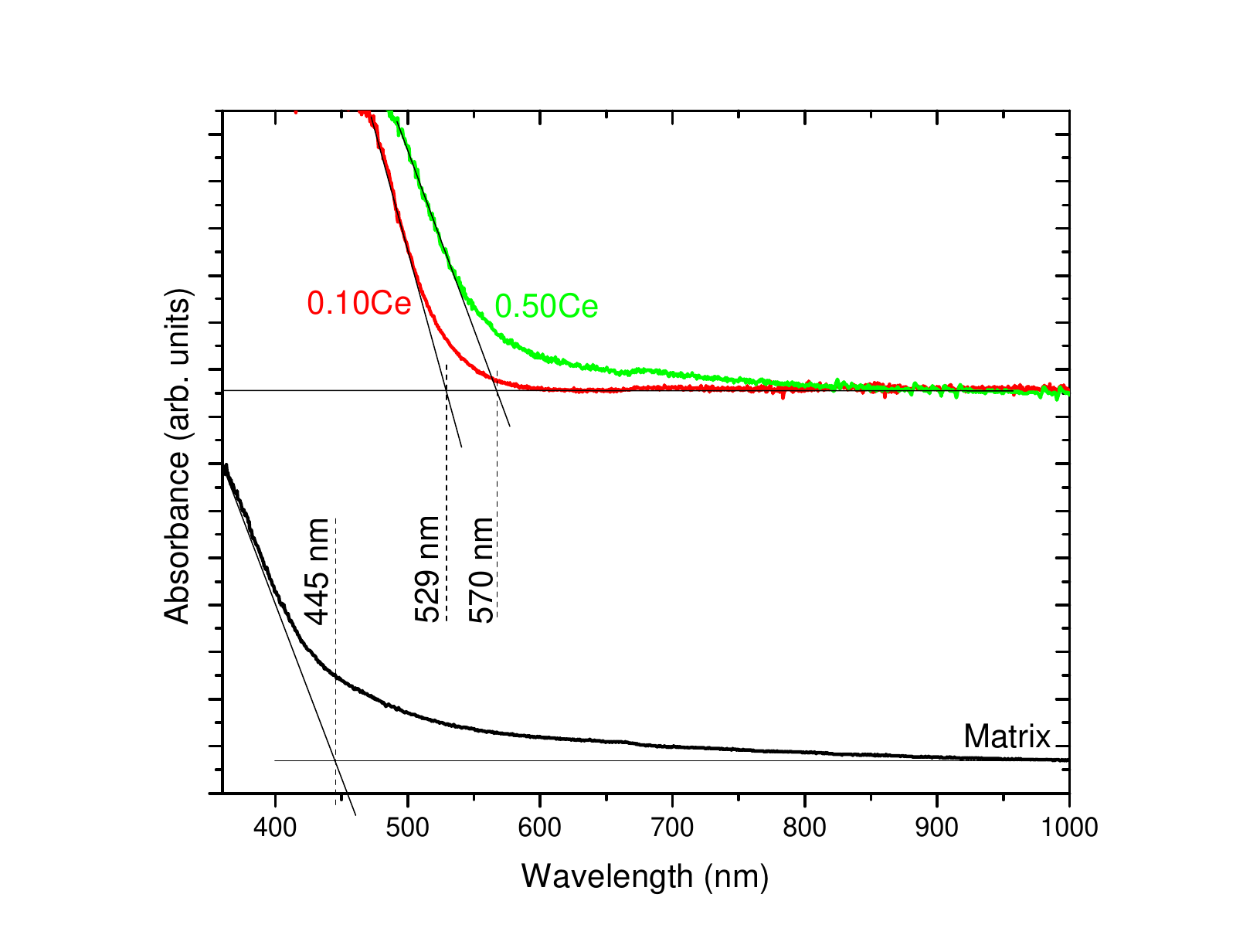}%
\caption{Absorbance spectra of TZNL samples.}
\label{abs}
\end{figure}

Optical absorbance spectra are shown in figure \ref{abs}, where we can see a band around 450 nm in our matrix similar to the observed by Costa et al \cite{Santos2018} in a binary lithium-tellurite glass, which is attributed to Te$^{4+}$ ions. The cut-off wavelengths were estimated as indicated in the figure and by adding Cerium its position has moved from 445 nm in the matrix to 529 nm and 570 nm in 0.10Ce and 0.50Ce samples, respectively. 

The literature on these glasses indicates a few different mechanisms to explain such redshift. The most common is the 4f-5d absorption of $Ce^{3+}$ \cite{Sasikala2012,Zheng2013,Tao2012,Huang2016a,Su2018}. However this same transition should result in intense visible emission from $Ce^{3+}$, and to the best of our knowledge, this has not been reported in any tellurite glass. Meanwhile, Ce$^{4+}$ may also introduce strong UV-VIS absorption due to CT transition \cite{Ebendorff-Heidepriem2000a, Zhou2015a}, but we found no reports in the literature mentioning the presence of this Cerium valence state in tellurite glasses. Another possibility often found to explain this redshift is the conversion of TeO$_3$ units into TeO$_{3+1}$ and TeO$_4$ units, as well the formation of more bridging oxygen (NBO) \cite{Biirger1992, Swapna2016, Elkhoshkhany2018}. A Fourier-transform infra-red (FTIR) study was performed and no significant change that could be related to the NBO's was detected. The occurrence of these modifications was already considered to explain our DSC results, though it is difficult to explain such a significant redshift of 125 nm (445 nm to 570 nm) with only this latter hypothesis. O$^{2-}$$\rightarrow$Ce$^{4+}$ CT is expected to be much more intense than O$^{2-}$$\rightarrow$Ce$^{3+}$ CT or 4f-5d absorption of Ce$^{3+}$ \cite{Zhao2018}. In this way Ce$^{4+}$ seems more likely to explain the redshift observed.

The following luminescence and excitation study provides some perspective about what could, or could not be the explanation for the redshift.

\subsection{Luminescence}
\begin{figure}[h]
\centering
\includegraphics[trim=1cm 1cm 2cm 1cm, clip,scale=0.35]{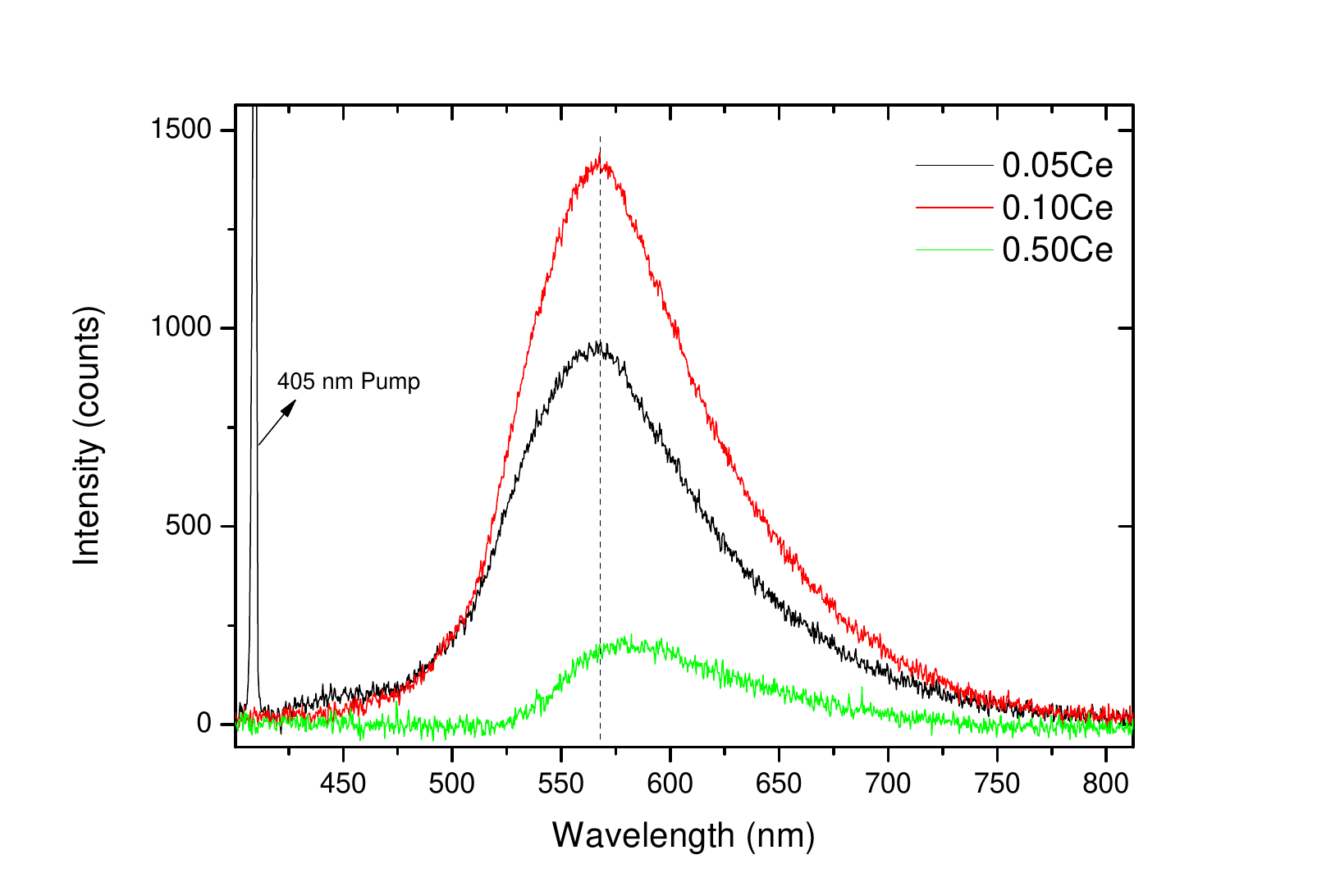}%
\caption{TZNL luminescence spectra under 405 nm excitation. The inset shows a picture of one of the samples.}
\label{luminescence}
\end{figure}

The luminescence spectra of the 0.10Ce and 0.50 Ce are shown in figure \ref{luminescence} under 405 nm laser excitation and in the inset a picture from the sample 0.10Ce demonstrating the green/yellow emission we have detected. The sample 0.05Ce was also included in this experiment, and we were not able to detect any luminescence from the matrix using this setup.

\begin{figure}[h]
\centering
\includegraphics[trim=2cm 1cm 2cm 1cm, clip,scale=0.37]{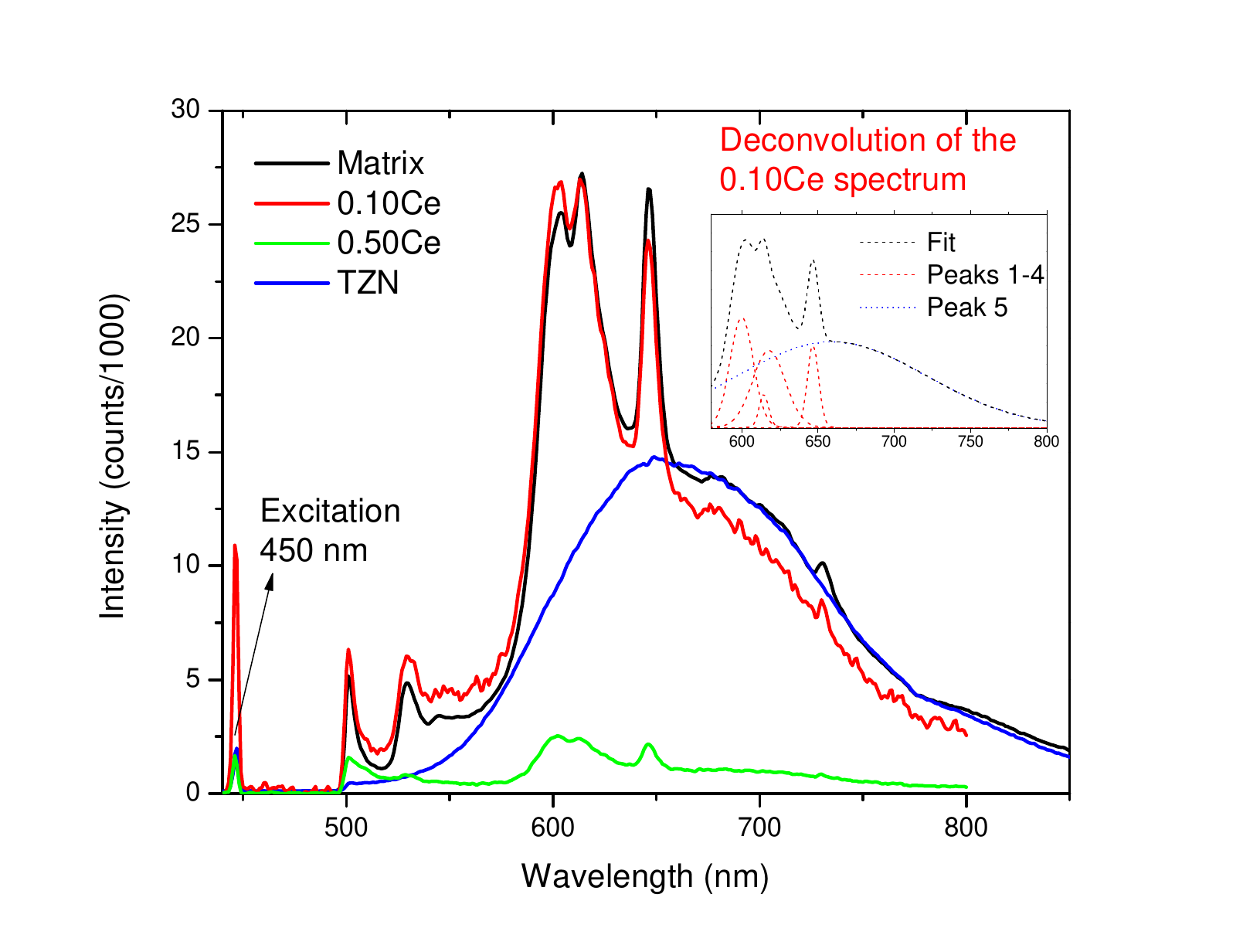}%
\caption{Luminescence spectra under 450 nm excitation, where a long pass filter (cut-on wavelength 500 nm) was used. The blue curve named TZN is from an additional sample prepared without Lanthanum oxide. The inset shows a deconvolution of the 0.10Ce spectrum by fitting it to five gaussians.}
\label{pr}
\end{figure}

These measurements were performed with different pieces of the samples, with the laser beam positioned near the border of the glass and the luminescence collected perpendicularly to the beam, in the lateral surface of the sample. The two surfaces were polished carefully, and in this way, the intensities that are shown in figure \ref{luminescence} can be compared. At first glance, we believed this luminescence was from Ce$^{3+}$ emission due to the decays $^2D_{3/2}$:$^2F_{7/2}$ $^2D_{3/2}$:$^2F_{5/2}$. The distance between these two transitions is about 2250 $cm^{-1}$ \cite{Zhang2015b} and it often results in a characteristic broad luminescence \cite{Silveira2012a} in the visible that is similar to the spectra we have shown in figure \ref{luminescence}. However, by using another setup where more pumping power was available we measured the luminescence under $450$ nm pumping, including a rare-earth free sample (TZN) as one can see in figure \ref{pr}, where the inset shows a deconvolution of the bands in the range 580-800 nm by a multi-gaussian fit.
 
A broad emission centered at about 650 nm was observed, as well as some unexpected emission lines which fit perfectly well to several Pr$^{3+}$ transitions. Traces of $Pr^{3+}$ can be found in both lanthanum and cerium oxide, and this ion has intense absorption at $\sim$ 445 nm (very near the pumping wavelength), $\sim$ 470 nm and $\sim$ 485 nm \cite{Belancon2014c, Taniguchi2018a} due to the ground state absorption of the $^3P_{2}$, [$^3P_{1},^1I_{6}$] and $^3P_{0}$ levels, respectively. In this way, pumping at 450 nm we are favouring the population of these levels, which by consequence results in the $Pr^{3+}$ emission lines near 500 nm ($^3P_{0}\rightarrow$ $^3H_{4}$ or $^3P_{2}\rightarrow$ $^3H_{5}$); around 530 nm ($^3P_{1}\rightarrow$ $^3H_{5}$); around 600 nm ($^1D_{2}\rightarrow$ $^3H_{4}$); around 615 nm ($^3P_{0}\rightarrow$ $^3H_{6}$) and at 645 nm ($^3P_{0}\rightarrow$ $^3F_{2,3}$). For the sample 0.50Ce the luminescence intensity has diminished significantly.

We prepared the additional sample, denoted by TZN (Tellurium, Zinc, Sodium) to confirm the origin of these emission lines. As one can see in figure \ref{pr}, TZN exhibits the same broad emission around 650 nm, though the emission lines are absent, which we believe is enough to confirm that the broadband is not related to Cerium or other rare-earth.

The only explanation for the origin of this broadband emission around 650 nm is the Te$^{4+}$ ions. Very similar luminescence observations attributed to this ion can be found in the literature \cite{Santos2018}, which also agrees with the absorption band around 450 nm found in our matrix sample. In this way, the remaining question is about the origin of the broad luminescence around 570 nm shown in figure \ref{luminescence}. To answer that we performed the optical excitation study presented next.

\subsection{Excitation}

\begin{figure}[h]
\centering
\includegraphics[trim=2cm 1cm 2cm 1cm, clip,scale=0.35]{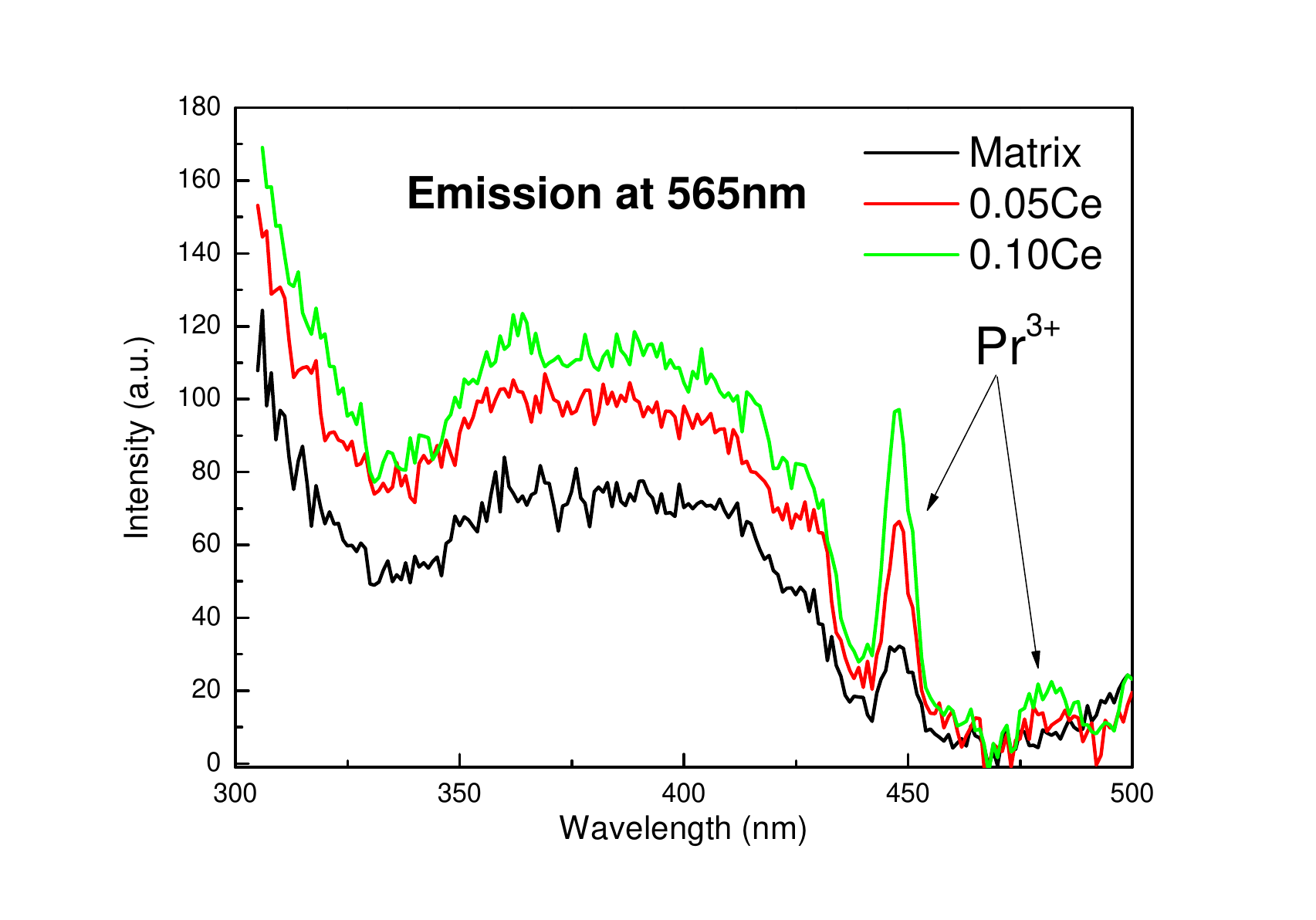}%
\caption{Excitation spectra for the emission band around $\sim 565$ nm for studied samples, including the matrix. As one can see, Cerium Oxide has enhanced slightly the intensity, though it did not introduced a new band.}
\label{excit}
\end{figure}

In figure \ref{excit} we can see the excitation spectra for the emission band at $\sim570$ nm measured by a photomultiplier and using a lock-in amplifier. This setup provided higher sensitivity and confirmed that all samples, including the matrix, have similar excitation bands, though they seem a bit more intense for the sample 0.10Ce.

As one can see, besides a broad excitation band between 350-430 nm some excitations lines can also be observed. All of these lines can be explained by considering the presence of Pr$^{3+}$ traces, which also agrees with the luminescence spectra shown in figure \ref{pr}. The sample highly doped (0.10Ce) should also have more Pr$^{3+}$ impurities, which is confirmed by the more intense excitation at 445 nm and the band around 480 nm, as we can see in figure \ref{excit}. As mentioned before, Pr$^{3+}$ has ground state absorption due to the $^3P_{2}$ (445 nm), [$^3P_{1},^1I_{6}$] (470 nm) and $^3P_{0}$ (480 nm) levels, which fits perfectly well to the observations made here. We also tried to measure the intensity decays employing the same setup used to acquire the excitation spectra, but unfortunately, the signal was too weak and short to be measured at our experimental conditions.

Besides the Pr$^{3+}$ related bands mentioned above, the shapes of the excitation spectra for all samples are identical, i. e.  Ce$^{3+}$ bands were not detected and the broad luminescence in the visible range we have observed is not related to Ce$^{3+}$. In this way, we believe the redshift observed in the absorption spectra of our samples could not be attributed to this valence state of Cerium.

\section{Discussion}

The DSC and XRD results does not indicates any important structural difference among our samples, i. e. glass transition temperature as well the amorphous nature has been maintained as Cerium Oxide concentration increased. Though it could be possible that Cerium oxide is modifying the glass network and transforming some tellurium units.

Concerning the spectroscopic properties, the redshift in the absorption edge can not be explained by the presence of Ce$^{3+}$, once that we have not found any signal of this ion in the excitation measurements we have performed. Even though the luminescence observed in figure \ref{luminescence} decreases for the sample 0.50Ce, what could be interpreted as concentration quenching of the Ce$^{3+}$, one should remember that the cut-off wavelength for this highly doped sample is at 570 nm. O$^{2-}{\rightarrow}$Ce$^{4+}$ CT, which does not result in radiative emissions and could be the reason for the diminishing luminescence observed.

The excitation spectra does not match the absorption spectra, and by this way as Cerium oxide concentration increases these results are indicating that Ce$^{4+}$ is the responsible for the redshift of the absorption edge. Though the conversion of Tellurium units cannot be discarded and the dopant can be modifying the glass network and favouring the formation of more ionic Te$^{4+}$ active ions. This interpretation explains the fact that the broadband emissions are found even in the TZN (rare-earth free) sample, and their intensities increases for the sample 0.05Ce and 0.10Ce, while O$^{2-}{\rightarrow}$Ce$^{4+}$ CT dominates the absorption spectra below 570 nm for the sample 0.50Ce. 

The overlap between Ce$^{3+}$ and Ce$^{4+}$ absorption bands is common in glasses \cite{Stroud1961a,Lin1995,Ebendorff-Heidepriem2000a,Pinet2006,Zhao2018}, and in our case the samples were prepared in air, which favors the formation of Ce$^{4+}$. Ce$^{3+}$, if present, would transfer energy to the more abundant Ce$^{4+}$ which is an ``efficient killer'' \cite{Lin1995} for the Ce$^{3+}$ emission.

In this sense, the broad excitation band between 350-430 nm observed in figure \ref{excit} can be explained by direct excitation of the Te$^{4+}$ ions, a result similar to others already reported for tellurite glasses \cite{Santos2018,HumbertodaCunhaAndrade2018}.

\section{Conclusion}

The results reported in this work have demonstrated that cerium oxide incorporation in zinc-tellurite glasses, for samples prepared in air, would result in Ce$^{4+}$ rich samples. Even though the presence of this ion in cerium doped tellurite glasses are barely reported, the abundance of the oxidized valence state could explain why emission of Ce$^{3+}$ has not been reported in tellurites, even though many authors confirmed that this ion was present in their tellurite samples.

The broadband luminescence in the visible, around 570 nm, and 650 nm under excitation at 405 nm and 450 nm reported in this work are due Te$^{4+}$. The population of this ion seems to increase with cerium oxide concentration, which acts modifying the glass network as suggested by our DSC results, though at the same time the dopant is introducing Ce$^{4+}$ that will end killing the radiative emissions. If present, Ce$^{3+}$ should be very diluted and its 4f-5d transition does not play any role in the visible luminescence we have observed.

In conclusion, our work confirms that cerium oxide doped TZNL glasses with good thermal stability could be achieved under the conventional melting quenching process under air. Te$^{4+}$ active centers exhibiting an interesting emission are present, though to develop spectral converters for solar cells or other devices such emission intensity has to be enhanced. The incorporation of cerium oxide increased the absorption of light, though it only slightly contributed to the broadband visible emission from Te$^{4+}$ for dopant concentration up to 0.1\%mol. Further studies are needed to verify if the population of the reduced valence state of Cerium can be increased by modifying the glass composition and the fabrication process, as demonstrated in other glasses \cite{Zhao2018}. Though Ce$^{4+}$ may be an excellent UV blocking agent, the prevalence of the reduced valence state would be more useful for other applications. These are some of the challenges we will investigate in future work.

\label{S:4}

\section{Acknowledgments}
\label{S:5}
The authors would like to thank Brazilian agency CNPq (grant $480576/2013-0$) and CAPES for their financial support, and to the ``Laboratório Central de Análises''.






\bibliographystyle{model1-num-names}

\end{document}